\documentstyle[epsfig,aps,floats,preprint,tighten]{revtex}

\def\beq{\begin{equation}}
\def\eeq{\end{equation}}
\def\qvec{{\bf q}}
\def\pbar{\overline{\bf p}}
\def\qhat{\hat{\bf q}}
\def\etal{{\it et al.}}
\def\epsbol{\mbox{\boldmath$\epsilon$}}
\def\sigbol{\mbox{\boldmath$\sigma$}}
\def\mpi{M_\pi}
\def\mn{m_{\scriptscriptstyle N}}
\def\N{{\scriptscriptstyle N}}
\def\A{{\scriptscriptstyle A}}

\def\degree{$^{\rm o}$\ }
\def\half{{\textstyle {1\over 2}}}
\def\t{{\bf t}}
\def\smfrac#1#2{{\textstyle {#1\over #2}}}

\def\insertfigone{
\begin{figure}
  \begin{center} \mbox{\epsfig{file=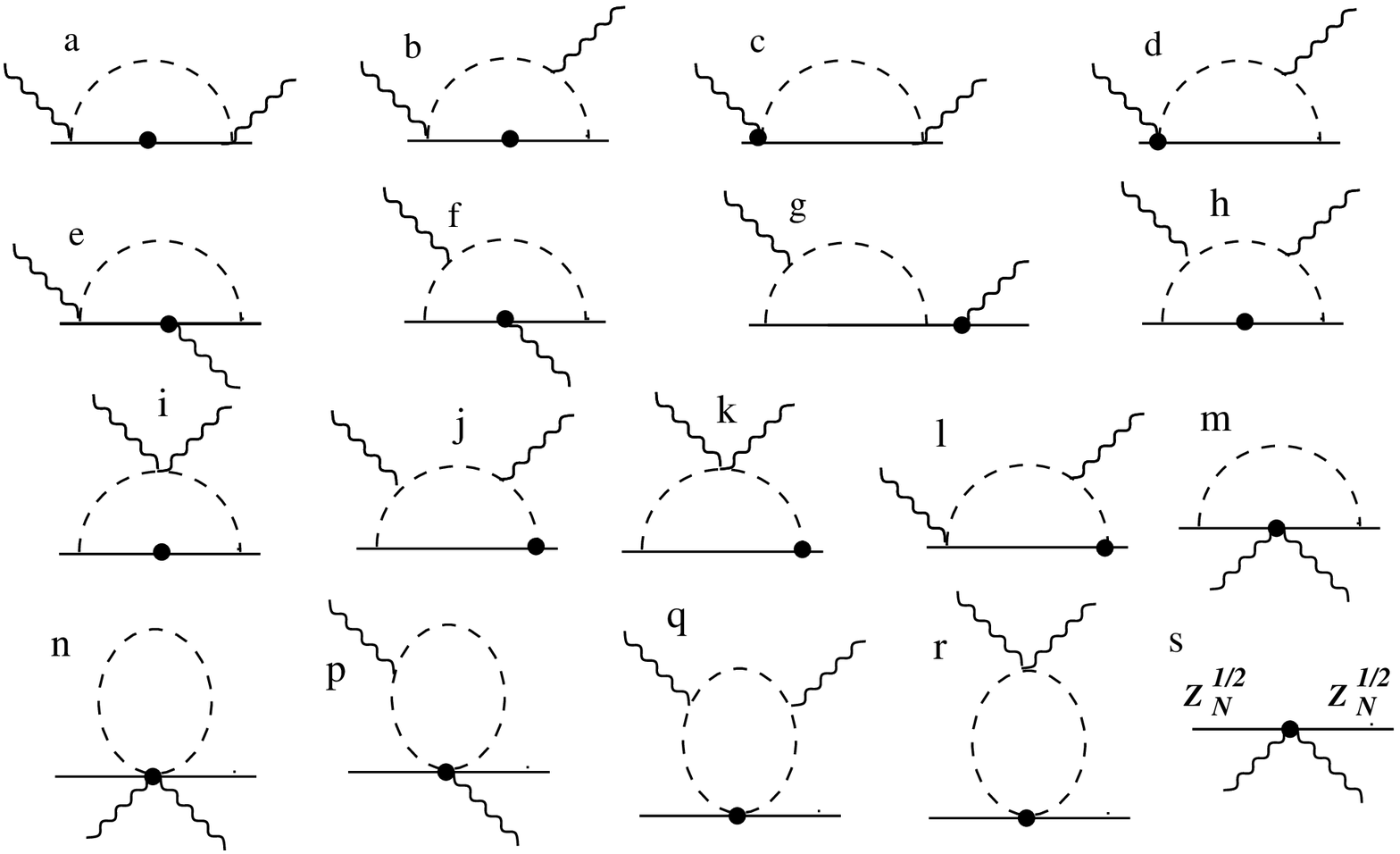,width=12truecm,angle=0}}
  \end{center}
{\bf Fig.~1:}
Diagrams which contribute to spin-dependent forward Compton
scattering in the $\epsilon\cdot v=0$ gauge at 4th order. The solid dots 
are vertices from ${\cal L}^{(2)}$.
\end{figure}}

\def\insertfigtwo{
\begin{figure}
 \begin{center} \mbox{\epsfig{file=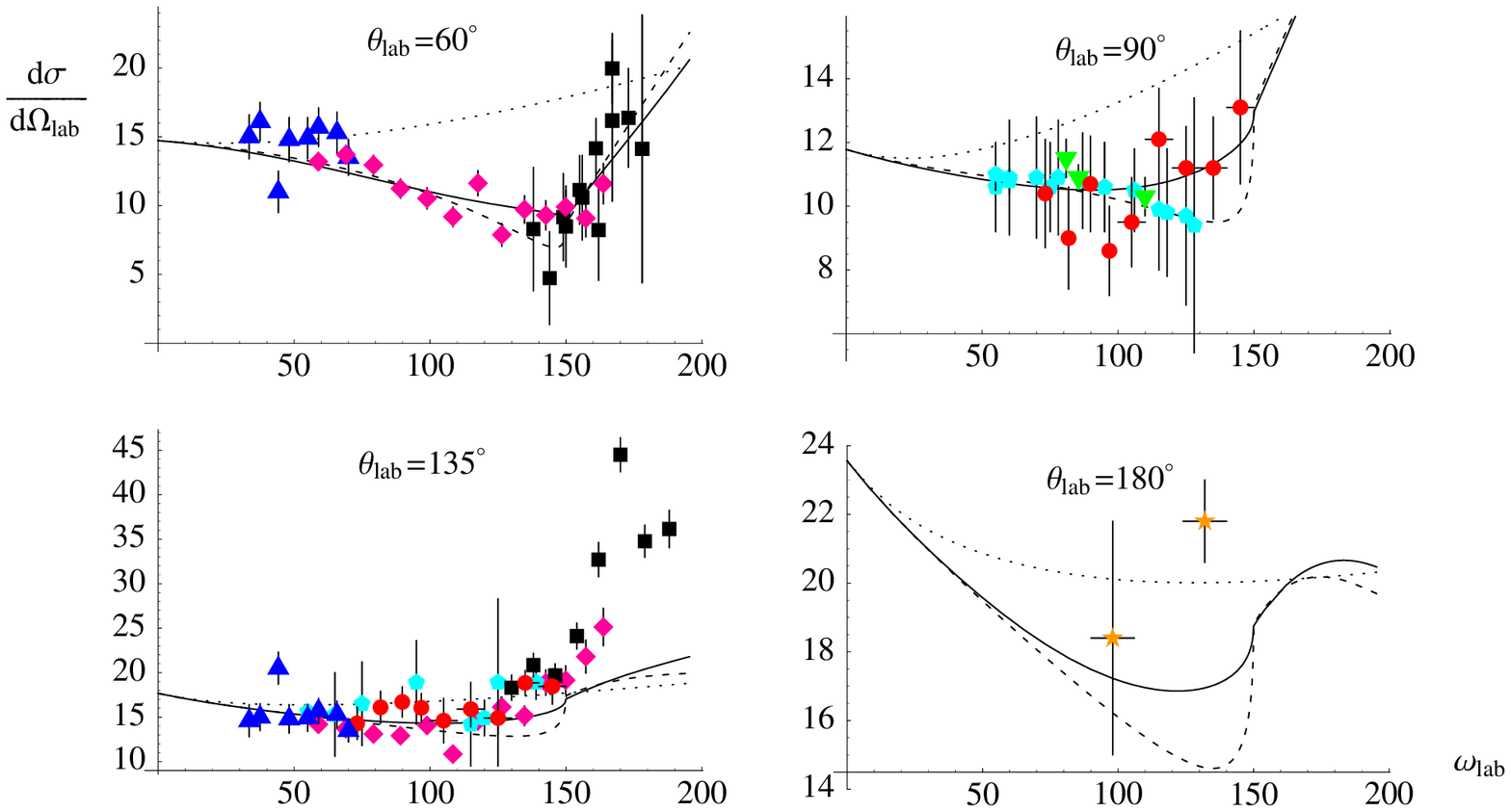,width=16truecm,angle=0}}
  \end{center}
\vskip-0.8cm {\bf Fig.~2:}
The fourth order (solid line), third order (dashes) and Born (dotted)
predictions for various lab angles.  
The data are from Baranov, Moscow 
(green inverted triangles) \cite{baranov}, Ziegler, Mainz (orange stars)
\cite{ziegler}, Federspiel, Illinois (blue triangles) \cite{federspiel,baranov2},
Hallin, Saskatoon (black squares) \cite{hallin}, MacGibbon, Saskatoon
(red circles) \cite{macgibbon}, Olmos de Le\'on, Mainz (magenta diamonds) \cite{deLeon}
and various earlier experiments, listed by Baranov (cyan pentagons) \cite{baranov2}. 
Cross sections are in  units of nb/sr, and energies in MeV.  Data points have been included
if their angle is within 2 degrees of the quoted value.
\end{figure}}

\def\insertfigthree{
\begin{figure}
  \begin{center} \mbox{\epsfig{file=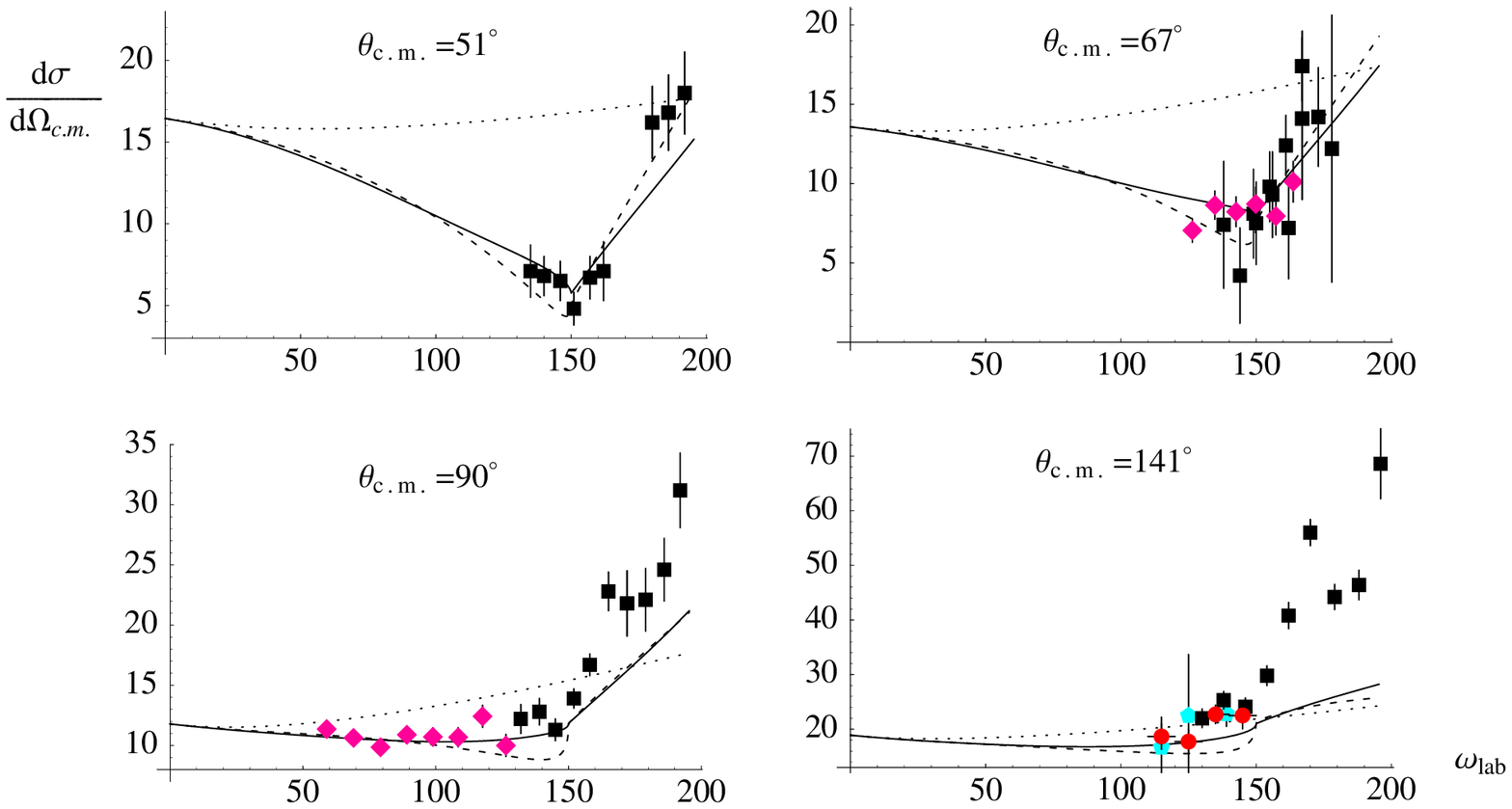,width=16truecm,angle=0}}
  \end{center}
\vskip-0.8cm {\bf Fig.~3:}
Predictions for various CM angles. See Fig.~1 for legend.
\end{figure}}

\def\insertfigfour{
\begin{figure}
 \begin{center}
  \mbox{\epsfig{file=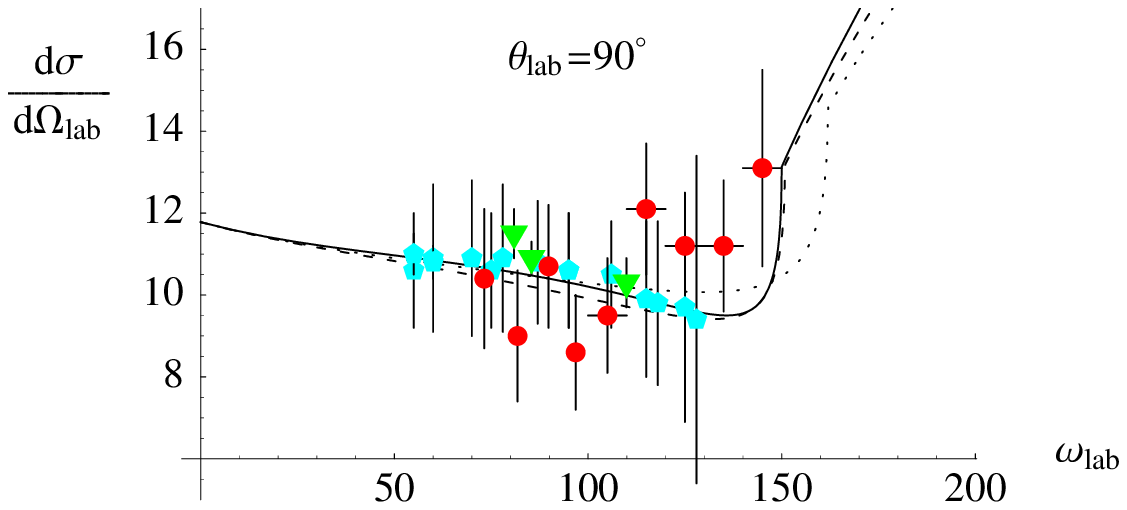,width=16truecm,angle=0}}
  \end{center} 
  \vskip-0.5cm   {\bf  Fig.~4:}  Effect   of  different   handling  of
  third-order  predictions.  Solid: prediction  taken as  referring to
  Breit  frame and  variable  change performed  to reproduce  the physical
  value  of the threshold;  dashed: as  solid, but  no variable  change; 
  dotted: predictions  taken as  referring  to the centre of  mass  frame, 
  as  in ref.~\cite{bab97}
\end{figure}}

\def\insertfigfive{
\begin{figure}
 \begin{center} \mbox{\epsfig{file=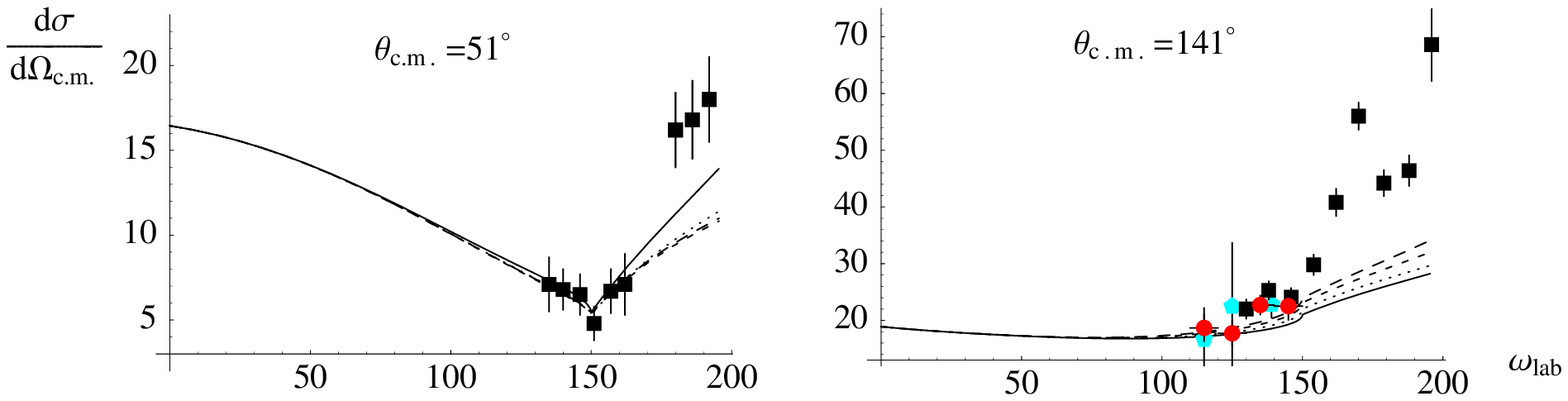,width=16truecm,angle=0}}
  \end{center}
\vskip-0.8cm {\bf Fig.~5:}
Effect of altering the spin polarisabilities by hand. Solid line: 
4th order HBCPT (as in 
previous figures); long dashes, short dashes and dots: polarisabilities taken
from the DR analyses of refs~\cite{mainz1,mainz2,BGLMN} respectively.
\end{figure}}

\def\insertfigsix{
\begin{figure}
 \begin{center} \mbox{\epsfig{file=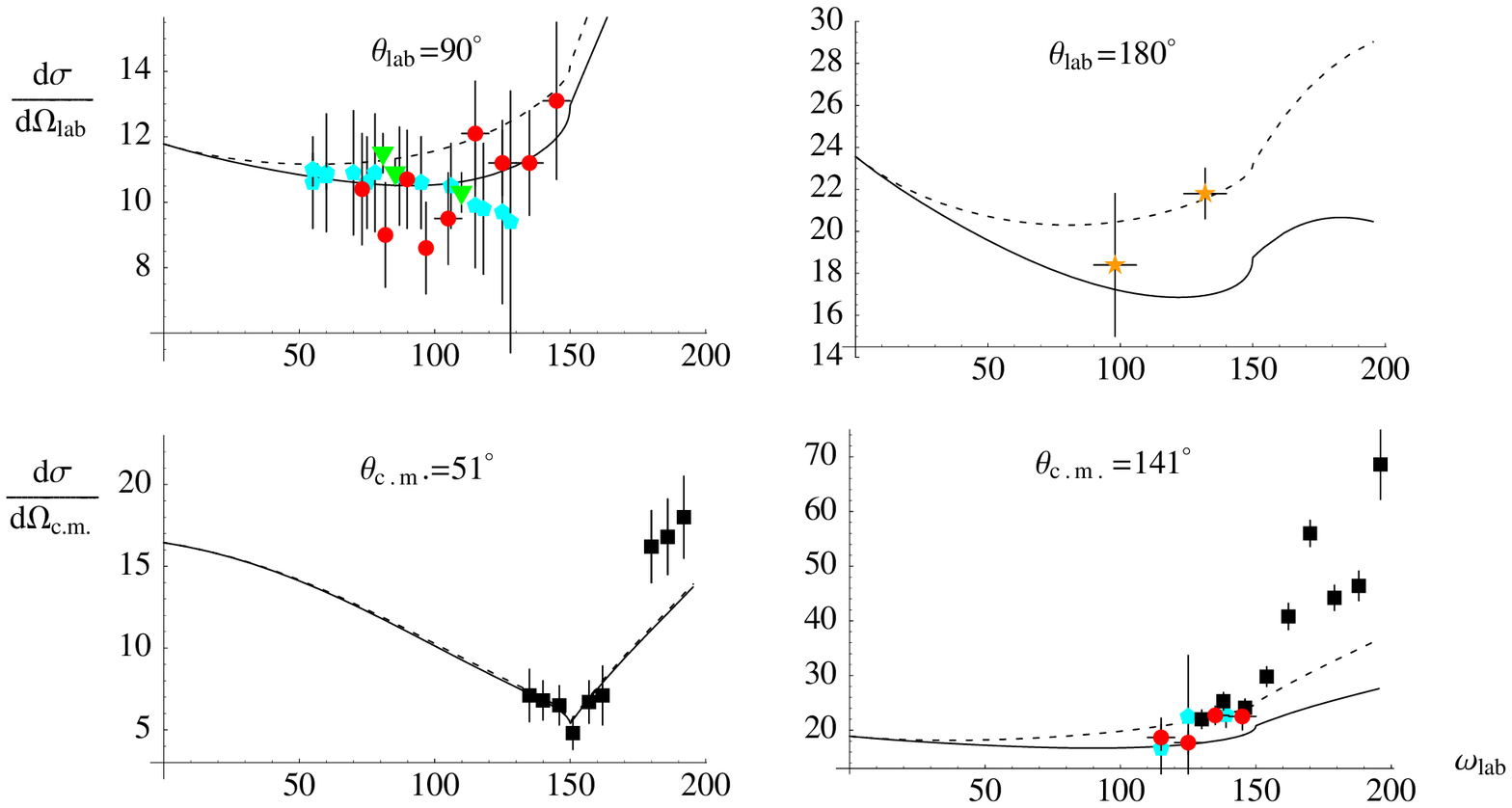,width=16truecm,angle=0}}
  \end{center}
\vskip-0.8cm {\bf Fig.~6:}
Fourth order HBCPT with $\alpha$ and $\beta$ set to their PDG values (solid),
and to $\alpha=8$, $\beta=6  \times   10^{-4}$~fm$^3$  (dashed).
\end{figure}}

\def\insertfigseven{
\begin{figure}
 \begin{center}
  \mbox{\epsfig{file=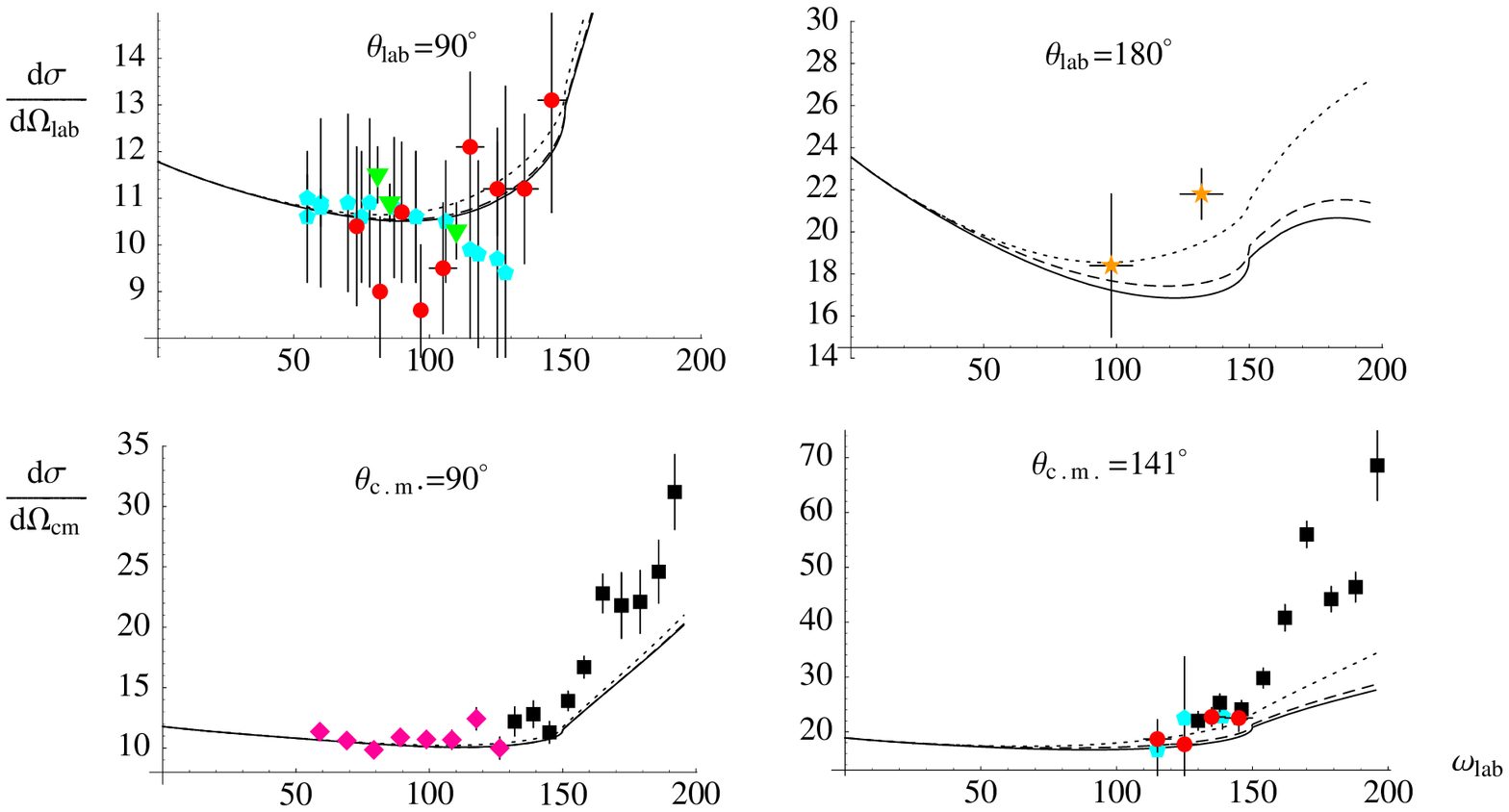,width=16truecm,angle=0}}
  \end{center}  
  \vskip-0.8cm {\bf Fig.~7:}  Effect of  varying  the pion-scattering
  LEC's  $c_i$.   Values  are  taken   from  Fettes  and   Mei\ss  ner
  \cite{fettes00}. Solid  and dashed lines  are the second  order fits
  based on KA85 and SP98; dotted  is the fourth order fit to SP98 from
  Table 1 of that work.
\end{figure}}

\begin{document}
\draft

\title{Compton scattering from the proton at NLO in the chiral expansion}
\author{
Judith A. McGovern\footnote{Electronic address: judith.mcgovern@man.ac.uk} }
\vskip 20pt
\address{Theoretical Physics Group, Department of Physics and Astronomy\\
University of Manchester, Manchester, M13 9PL, U.K.}
\nopagebreak
\maketitle

\begin{abstract}
We  present calculations  of differential  cross sections  for Compton
scattering  from the  proton,  using amplitudes  calculated to  fourth
order  in heavy baryon  chiral perturbation  theory.  We  compare with
available data up  to 200~MeV.  We find that  the agreement for angles
below 90\degree is acceptable over the whole energy range, but that at
more backward angles the  agreement decreases above about 100~MeV, and
fails completely above the photoproduction threshold.
\end{abstract}


\section{Introduction}

Heavy  baryon  chiral  perturbation  theory (HBCPT)  is  a  systematic
framework in which to describe the interactions of nucleons with pions
and photons.   Respecting gauge and Lorentz  invariance, it reproduces
low-energy theorems, which have been known since the 50's and 60's, as
the first  terms in chiral expansions,  and in a number  of cases also
makes  further parameter-free  predictions.  One  of the  simplest and
cleanest processes to which it can be applied, theoretically at least,
is Compton scattering,  and indeed this was first  studied in the very
early  days  of the  theory.   First  spin-independent scattering  was
considered.  At  lowest order (technically second order  in HBCPT) the
scattering  amplitude is simply  the energy-independent  Thomson term,
which  is  entirely  independent   of  all  details  of  the  proton's
structure.  This  arises simply as  a seagull term in  the Lagrangian,
one of  the terms that arise  when antinucleons are  integrated out of
the lowest-order relativistic Lagrangian.  Energy and angle dependence
in the amplitude  enter at third order, where  the only diagrams which
contribute  are one-pion  loop graphs.   The next  lowest terms  in an
energy  expansion   of  the  amplitude  (of   order  $\omega^2$)  have
coefficient  which  are  called  the  polarisabilities,  $\alpha$  and
$\beta$, and at this order in HBCPT they are simply predicted in terms
of well-known nucleon and pion properties \cite{ber92}.  The numerical
values, $\alpha=12.5 \times 10^{-4}$~fm$^3$ and $\beta=\alpha/10$, are
in remarkably good  agreement with the current world  averages for the
proton,    $\alpha=12.1\pm0.8\pm0.5    \times   10^{-4}$~fm$^3$    and
$\alpha+\beta=14.1\pm0.5 \times  10^{-4}$~fm$^3$ \cite{pdg}.  This was
an early triumph for HBCPT.

The calculation  of the forward  spin-independent scattering amplitude
to fourth order followed quickly \cite{ber93,ber93a}.  However at that
order there  is no further  predictive power for  the polarisabilities
within the strict  framework of the theory, as  there are four unknown
low-energy  constants  in   the  fourth-order  Lagrangian  which  give
contributions   to   the   isoscalar   and  isovector   $\alpha$   and
$\beta$. Also, the amplitude is not easily compared with experiment on
its own.

The next step was the calculation of the full spin-dependent amplitude
to  third order\cite{ber95}.  There  are four  independent amplitudes,
each  with a  leading term  (of order  $\omega$) which  is given  by a
low-energy  theorem   \cite{LGG}.  The   next  terms,  of   ${\cal  O}
(\omega^3)$, have  coefficients, denoted $\gamma_i$,  which are called
spin  polarisabilities.   Unfortunately   these  are  not  well  known
experimentally;  indeed  till  recently  only the  contribution  which
enters forward  scattering, called $\gamma_0$,  was known at  all, and
that only  from pion photoproduction  data rather than  direct Compton
scattering    experiments\cite{karl,sand,drech98}.     More   recently
$\gamma_\pi$  (for  backward   scattering)  has  been  extracted  from
unpolarised Compton  scattering experiments, with  conflicting results
\cite{tonn,wiss}.    The  first   direct  measurements   of  polarised
photon-proton  scattering   at  MAMI  yield  a   more  reliable  value
$\gamma_0$ of $-0.8\pm0.1  \times 10^{-4}$~fm$^4$ \cite{pedroni} which
is  within  the  range   of  previous  estimates.   In  addition  pion
photoproduction data has been  used as input into fixed-$t$ dispersion
relations   to  obtain  estimates   of  the   $\gamma_i$  individually
\cite{mainz1,BGLMN,mainz2}.   Unfortunately  the  lowest  order  HBCPT
values are  in total disagreement  with the experimental  numbers, the
only saving  point of agreement  being that the  isovector quantities,
which vanish  in LO HBCPT, do seem  to be smaller in  general than the
isoscalar ones.

Most recently, three groups  have calculated the spin polarisabilities
to fourth order  (NLO) \cite{ji00,kumar99,gellas,kumar00}.  Unlike the
spin-independent case, this order contains  no LEC's, and so there are
again genuine predictions.  Surprisingly  the NLO contributions are as
large as  the LO  ones.  The agreement  with the  experimental numbers
however is variable.

Of course, classic HBCPT has no explicit $\Delta$.  This is assumed to
have  been integrated  out, and  its  legacy is  in the  LEC's of  the
nucleon-only  theory.   It  is  seen  first  at  different  orders  in
different processes:  second order  for pion scattering,  fourth order
for  spin-independent  and  fifth  order for  spin  dependent  Compton
scattering.   It is  expected  to  be important;  it  may improve  the
agreement  of  the  spin  polarisabilities,  but  risks  spoiling  the
spin-independent ones.

\insertfigone

All this emphasis on the  polarisabilities obscures the fact the HBCPT
does in fact give complete  amplitudes for scattering, which one might
hope would be believable up to energies of the order of the pion mass,
since the chiral  expansion is both in powers  of $\omega/\Lambda$ and
$\mpi/\Lambda$ (where $\Lambda$ is either $\mn$ or the chiral scale $4
\pi f_\pi$).  When the  full third-order amplitudes were published, it
was claimed that they gave  good agreement with cross section data for
unpolarised scattering up to the photoproduction threshold.  However a
more  wide-ranging study  was done  by Babusci  \etal\cite{bab97}, who
showed  that the agreement  started to  breakdown for  backward angles
above about 120~MeV.  They stopped their comparisons at the threshold.

We have now  calculated the full scattering amplitude  to fourth order
in  HBCPT.    Since  the   fourth-order  contributions  to   the  spin
polarisabilities  are so  large,  it  is clearly  of  interest to  see
whether  the  full amplitudes  and  cross  sections  are also  changed
dramatically, and this is the subject of this paper.  We have compared
with       the      available       data      up       to      200~MeV
\cite{baranov,ziegler,federspiel,hallin,macgibbon},  and find  that in
fact the difference between  third- and fourth-order cross sections is
small.  Above  the threshold  the  good  agreement  at forward  angles
continues,  while  the  lack   of  agreement  at  backward  angles  is
dramatically worse.

\section{The cross section in HBCPT}

The usual notation for the scattering amplitude in the Breit frame is,
for incoming real photons of  energy $\omega$ and momentum $\qvec$ and
outgoing real photons of the same energy and momentum $\qvec'$,

\insertfigtwo

\begin{eqnarray}
T&=&\epsilon'^\mu\Theta_{\mu\nu} \epsilon^\nu\nonumber\\
&=&\epsbol'\cdot\epsbol\,A_1(\omega,\theta) 
+\epsbol'\cdot\qhat\,\epsbol\cdot\qhat'\,A_2(\omega,\theta) \nonumber\\
&&+i\sigbol\cdot(\epsbol'\times\epsbol)\,A_3(\omega,\theta)+
i\sigbol\cdot(\qhat'\times \qhat)\,\epsbol'\cdot\epsbol \,
             A_4(\omega,\theta)\nonumber\\
&&+\Bigl(i\sigbol\cdot(\epsbol'\times \qhat)\,\epsbol\cdot\qhat'-
i\sigbol\cdot(\epsbol\times \qhat')\,\epsbol'\cdot\qhat\Bigr)\,
             A_5(\omega,\theta)\nonumber\\
&&+\Bigl(i\sigbol\cdot(\epsbol'\times \qhat')\,\epsbol\cdot\qhat'-
i\sigbol\cdot(\epsbol\times \qhat)\,\epsbol'\cdot\qhat\Bigr)\,
             A_6(\omega,\theta),
\label{amp}
\end{eqnarray}
where hats indicate unit vectors.  By crossing symmetry the functions
$A_i$  are even  in $\omega$  for $i=1,2$  and odd  for  $i=3-6$.  The
amplitudes may be decomposed into pole and non-pole pieces.  The pole,
which  occurs  at unphysical  values  of  $\omega$  and  $t\,
(=2\omega^2(\cos\theta-1))$,  arises  from   Born  diagrams  in  which  the
intermediate nucleon is on shell, and the contribution containing this
pole can be calculated using Dirac nucleons and
on-shell photon-nucleon couplings.   The leading  terms are  given by
low-energy  theorems\cite{LGG}.    With the pole terms  truncated at fourth
order (that is, at ${\cal O}(1/\mn^3)$), the amplitudes read

\insertfigthree

\begin{eqnarray}
A_1(\omega,\theta)\!&=&\!-{Q^2e^2\over\mn}+{e^2\over 4\mn^3}
\Bigl((Q+\kappa)^2(1+\cos\theta)-Q^2\Bigr)(1-\cos\theta)\,\omega^2\nonumber\\
&&+ 4\pi(\alpha+\cos\theta\,\beta)\,\omega^2+{\cal O}(\omega^4)\nonumber\\
A_2(\omega,\theta)\!&=&\!{e^2\over 4\mn^3}\kappa(2Q+\kappa)\cos\theta\,\omega^2
-4\pi\beta\,\omega^2+{\cal O}(\omega^4)\nonumber\\
A_3(\omega,\theta)\!&=&\!
{e^2 \omega\over 2\mn^2}\Bigl(Q(Q+2\kappa)-(Q+\kappa)^2 \cos\theta\Bigr)
+A_3^{\pi^0}
+4\pi\omega^3(\gamma_1+\gamma_5\cos\theta)+{\cal O}(\omega^5)\nonumber\\
A_4(\omega,\theta)\!&=&\! -{e^2\omega \over 2\mn^2 }(Q+\kappa)^2 
+A_4^{\pi^0}+4\pi\omega^3 \gamma_2 +{\cal O}(\omega^5)\nonumber\\
A_5(\omega,\theta)\!&=&\! {e^2 \omega\over 2\mn^2 }(Q+\kappa)^2 
+A_5^{\pi^0}+4\pi\omega^3\gamma_4 +{\cal O}(\omega^5)\nonumber\\
A_6(\omega,\theta)\!&=&\! -{e^2 \omega\over 2\mn^2 }Q(Q+\kappa)
+A_6^{\pi^0}+4\pi\omega^3\gamma_3 +{\cal O}(\omega^5)
\label{ampexp}
\end{eqnarray}
where the charge of the  nucleon is $Q=(1+\tau_3)/2$ and its anomalous
magnetic   moment    is   $\kappa=(\kappa_s+\kappa_v\tau_3)/2$.  
The polarisability  terms are  the 
leading  terms  in the expansion  of  the   full  non-pole  amplitudes;  
the  polarisabilities  are  isospin dependent.  The
contribution  from the  $\pi^0$  graph, $A_i^{\pi^0}$,  has   been
separated  out,  though its  contribution  is often  included in  the
definition   of  the   polarisabilities.   Only   four  of   the  spin
polarisabilities   are  independent   since  three   are   related  by
$\gamma_5+\gamma_2+2\gamma_4=0$.    The  spin-dependent
amplitudes have no  pole contributions beyond the leading  ones at this
order. (At fifth  order there  are pieces  of  order $\omega^3/\mn^4$,
though  these   vanish  for  both  forward   and  backward  scattering
amplitudes.) The amplitudes are given for a nucleon spinor normalisation
of $\bar u u=1$.

\insertfigfour

To calculate these  amplitudes in HBCPT, we need  the all the diagrams
which appear  at fourth  order. (The pion-loop  diagrams are  shown in
Fig.~1.)  There is no clear  separation at a diagramatic level between
pole and non-pole; heavy-baryon seagulls give pieces of the Dirac pole
but, at  fourth order, they  also contribute to $\alpha$  and $\beta$.
Conversely  fourth-order loop  diagrams have  contributions  at ${\cal
O}(\omega)$ which  correct the bare  anomalous magnetic moment  in the
LET terms.  However, taking all diagrams together HBCPT reproduces the
pole terms of Eq.~\ref{ampexp}, with the physical (as opposed to bare)
nucleon  mass  and magnetic  moment,  as  an  expansion in  powers  of
$1/\mn$.  Everything  else is  non-pole.  (While this  separation into
pole  and non-pole  is useful  in  discussion, nothing  in this  paper
depends on them.   We simply calculate the full  amplitudes to a given
order and use them to produce cross sections.  This is not affected by
the divergence  of opinion  that is found  in the literature  over the
definition of the polarisabilities \cite{birse00}.)

For the  spin-dependent amplitudes,  no LEC's enter  except well-known
nucleon  and pion  properties.  We  take  $g_\A=1.267$, $f_\pi=92.32$,
$\kappa_s=-0.120$, $\kappa_v=3.706$ and $\mpi=139.6$ (as the loops are
of  charged  pions in  the  main).   However  in the  spin-independent
amplitudes two  sets of  LEC's enter.   One set is  the pair  of LEC's
which contribute  to $\alpha$  and $\beta$; these  have been  fixed by
requiring  the  full polarisabilities  (with  third  and fourth  order
contributions) to take the experimental values.  The other set are the
parameters  $c_i$ from ${\cal  L}^{(2)}$, which  enter in  the tadpole
graphs.   These are rather  poorly known;  furthermore they  have been
determined from  pion-nucleon scattering  at second, third  and fourth
order  with quite different  results (and  indeed from  different data
sets with different results).  Clearly the second-order determinations
are  the most  appropriate  to  use in  these  calculations as  vertex
dressing  is absent  at  this order.   We  have used  $c_1=-0.81$~GeV,
$c_2=2.5$~GeV  and  $c_3=-3.8$~GeV  \cite{fettes00}.   Sensitivity  to
these choices is explored later.

All our amplitudes are calculated in the Breit frame.  At third order,
the result  is the same in any  frame (so long as  the energies remain
low);  the   amplitudes  could   be  in  the   Breit  frame,   or  the
centre-of-mass frame, or even the lab frame (since $\omega-\omega'$ is
of higher chiral  order).  At fourth order however  the result will be
different  in  different  frames.   Interestingly there  is  still  no
dependence  on   $\omega-\omega'$,  but  there   will  be  expressions
involving  the average  nucleon three-momentum  $\pbar$. In  the Breit
frame  $\pbar$  vanishes, so  the  amplitudes  take  on by  far  their
simplest form.  In the centre-of-mass  frame $\pbar$ can be written in
terms  of the  photon  momenta,  and the  structures  which enter  are
unchanged,  but the  amplitudes  will have  terms  with the  ``wrong''
symmetry in $\omega$.   (These of course can be  obtained from a boost
of lower order amplitudes, and are not independent.)  The differential
cross section in any frame is a kinematic prefactor $\Phi^2$ times the
invariant  $|T|^2$, which in  terms of  the Breit-frame  amplitudes is
given by

\insertfigfive

\begin{eqnarray}
|T|^2&=& \half A_1^2  ( 1 + \cos^2\theta) + \half A_3^2
(3-\cos^2\theta) \nonumber \\ 
&& + \omega^2 \sin^2 \theta \, \bigl[ 4 A_3 A_6  +( A_3 A_4 +
2 A_3 A_5 -A_1 A_2)\cos \theta\bigr]\nonumber \\ 
&& + \omega^4 \sin^2 \theta \,\bigl[\half
A_2^2 \sin^2 \theta + \half A_4^2(1+\cos^2 \theta) + A_5^2
(1+2\cos^2 \theta) +3A_6^2 \nonumber \\ 
&& \kern2cm+ 2A_6 (A_4 + 3 A_5)\cos \theta + 2 A_4 A_5
\cos^2 \theta\bigr]
\label{tsqrd}
\end{eqnarray}
Above the threshold one should replace $A_i^2$ by $ |A_i|^2$ and 
$A_i A_j$ by $ \Re(A_i^* A_j)$.

Experimental  data,  though taken  in  the  lab  frame, are  sometimes
presented as centre-of-mass frame differential cross sections, so both
prefactors are needed:
\beq \Phi_{\rm lab}={1\over 4\pi}{\omega'\over
\omega};\qquad\qquad  \Phi_{\rm  cm}={1\over  4\pi}{\mn\over\sqrt{s}}.
\eeq

There is a further complication which stems from the heavy baryon
reduction.  Physically, there is a threshold at a lab energy of
$\omega_{\rm th}=\mpi(1+\mpi/2\mn)$.  However the HBCPT amplitudes at
third order have cusps at $\omega=\mpi$, and at fourth order they are
actually singular at this point.  The resolution was given by Bernard
\etal\ when discussing forward scattering \cite{ber93a}: if calculated
to all orders, the sum of all the singular terms will be precisely
what is required to shift the threshold to the physical value.  Thus
at finite order, to have the threshold at the right point and maintain
finite amplitudes, one changes variables from the Breit frame energy
$\omega$ to $\zeta\equiv\omega/\omega_{\rm th}$, then Taylor expands
and discards pieces of higher order.  At fourth order, therefore, if a
naive loop amplitude is $f^{(3)}(\omega/\mpi)+f^{(4)}(\omega/\mpi)$,
the shifted amplitude is 
\beq f^{(3)}(\zeta)+f^{(4)}(\zeta)
+\zeta\left({\omega_{\rm
th}\over\mpi}-1\right)_{\!\!\scriptscriptstyle LO} {d
f^{(3)}(\zeta)\over d\zeta}.  \eeq
 The subscript ``LO" indicates the
dropping of terms of ${\cal O}(1/\mn^2)$.  The shifted amplitude is
finite at $\zeta=1$. 

\insertfigsix

Since $|T|^2$ is  an invariant, it should make  no difference in which
frame it is  calculated, and we use the Breit  frame as the amplitudes
are  simplest  there.  Of  course  in  HBCPT  the invariance  is  only
respected up to  the order in $1/\mn$ to which we  are working, and so
the results obtained by calculating the amplitudes in different frames
would not in fact be identical.

Although  we have  calculated  the amplitudes  consistently to  fourth
order, those amplitudes substituted in Eq.~\ref{tsqrd} will not give a
differential cross section which  is of consistent order.  Since $A_1$
starts  at  second order  with  the  Thomson  term, knowledge  of  the
amplitude to fourth order gives the cross section to sixth order.  The
fourth order parts of all the other $A_i$ are not actually needed, and
the seventh- and eighth-order parts of $|A_1|^2$ should be subtracted.
We do not  do this.  The various orders  of contributions are actually
of very similar magnitude, and if we try to remove higher order pieces
we may not  even be left with a positive cross  section.  Thus all the
cross sections we present are calculated with fourth-order amplitudes,
and our kinematic  prefactors and change of variable  from $\omega$ to
$\zeta$ and from one frame to another are done exactly.

\section{Results}

In Figs.~2  and 3 we show  differential cross sections  at various lab
and centre-of-mass angles for which data exist below 200~MeV.  In each
graph we show the Born  (pion and nucleon pole) contribution at fourth
order, and the  HBCPT results with the amplitudes  calculated to third
and fourth order.  The third as well the fourth order results have had
the   variable   change   $\omega/\mpi\to\omega/\omega_{\rm  th}$   as
explained above; otherwise the threshold  would be in the wrong place.
This, and the fact that we  take the basic third-order amplitude to be
in the Breit frame, rather than the centre-of-mass frame, accounts for
the slight  differences between our  third-order results and  those of
Babusci  \etal \cite{bab97}.   In  Fig.~4 we  show  three third  order
curves: amplitudes calculated  in the Breit frame and  shifted, in the
Breit frame  and unshifted, and  in the CM  frame and unshifted  as in
ref.~\cite{bab97}.  The differences between these are higher order.

\insertfigseven

For  scattering angles  up to  90\degree the  fit at  either  third or
fourth order is very good except at the highest energies, 
and much better than the  Born terms alone.  In
the cusp region  the fourth order is arguably better than the third order
at all angles. The good  fit to the data
above  threshold is  a feature  which  has not  been remarked  before.
However for larger angles both curves  start to fall below the data as
the threshold  is approached, and  fall dramatically short  above.  Of
course one does  not expect agreement to continue  indefinitely, as in
experiments the $\Delta$ resonance is  clearly seen, and this will not
be reproduced by HBCPT.  However it is not obvious why backward angles
should be  so much  worse than  forward angles.  As  the data  in that
region is  only from one  experiment \cite{hallin} one  might question
its accuracy, but in fact  it fits well with dispersion analyses which
incorporate the more numerous higher energy data.

At fifth order, the new diagrams which enter fall into several groups.
First, there  are two-loop graphs, which  one might well  expect to be
small.   Then, there  are  one-loop graphs  with  two insertions  from
${\cal L}^{(2)}$.   Some of these give higher  $1/\mn$ corrections and
these  again  ought  to  be  small  (some  indeed  have  already  been
incorporated in the change  of variables which corrects the threshold)
but some, such  as two insertions of the  anomalous isovector magnetic
moment,  may  be numerically  significant.   Then  there are  one-loop
graphs  with  one  insertion   from  ${\cal  L}^{(3)}$,  and  seagulls
contributing directly  to the polarisabilities.   It is in  these that
the  effects  of the  $\Delta$  will  show  up in  the  spin-dependent
amplitudes,  and these  contributions are  expected to  be anomalously
large.  We  have not calculated fifth-order loop  graphs.  However the
seagulls may  be incorporated effortlessly, since the  free LEC's will
just be fit to give the ``experimental'' values of the $\gamma_i$.  In
Fig.~5  we show the  effect of  setting the  $\gamma_i$ to  the values
found     in    three     fixed-$t$     dispersion-relation    studies
\cite{mainz1,BGLMN,mainz2}.  Below threshold the agreement is slightly
improved,  but  above  the  enhancement  at  backward  angles  is  not
sufficient to give agreement.

Finally, we show the effects  of varying the input parameters within a
purely fourth-order calculation. These are the places where the effect
of the $\Delta$ is already present; for instance the tadpole graphs of
Fig.~1n-r could  arise from integrating out  an intermediate $\Delta$.
These   graphs   vanish   for   forward  scattering   but   contribute
significantly to  the spin-independent amplitudes  at backward angles.
First we show the results  of varying $\alpha-\beta$ while keeping the
sum fixed.   This shows up only  at backward angles;  to reproduce the
Mainz  data  at  180\degree we  need  to  make  a dramatic  shift,  to
$\alpha=8$  and  $\beta=6   \times  10^{-4}$~fm$^3$.  (Rather  similar
numbers were found  by Grie\ss hammer and Rupak in  a study of Compton
scattering  from the  deuteron  in the  framework  of effective  field
theory  without dynamical  pions \cite{griess}.)   However  even these
extreme values do  not cure the problem above  the threshold.  Then we
show the effect of using a  very different parameter set for the $c_i$
that  enter the  tadpole  graphs, while  keeping the  polarisabilities
fixed.  (These graphs do of course contribute to the polarisabilities,
as  given  ref.~\cite{ber93}.   However  the fourth  order  LEC's  are
adjusted  to give  the  measured polarisabilities  whatever the  other
contributions  might be.)  These  parameters are  quite uncertain  and
differ substantially in different  fits; we show two second-order fits
and  one fourth-order \cite{fettes00}.  Again, the  only effect  is at
backward angles, and none of  the parameter sets gives agreement above
the threshold.

Thus the conclusion  of this fourth-order study is  similar to that of
the third-order study of Babusci \etal\ \cite{bab97}. For angles up to
90\degree, HBCPT without an explicit $\Delta$ does an excellent job of
fitting the experimental differential Compton scattering cross section
up to about 170~MeV.  However  it fails at larger angles for photon
energies  above about  120~MeV. Very  large  fifth-order contributions
would be needed to approach  the experimental data above the threshold
at these angles.

This paper  was partly  supported by  the UK EPSRC.   I would  like to
acknowledge  the  hospitality of  the  Institute  for Nuclear  Theory,
Seattle, where a substantial part of  the work was carried out, and to
thank Mike Birse and Daniel Phillips for useful conversations.

\appendix

\section{Fourth-order loop amplitude in the Breit frame}

The  full amplitudes  in  the Breit  frame  for the diagrams of Fig.~1 are  as
follows.  The notation $\t_i$ is  used for the tensor structures which
multiply   the   amplitudes  $A_i$   of   Eq.~\ref{amp};  for   example
$\t_1=\epsbol'\cdot\epsbol$.

\begin{eqnarray}
T_a&=&{g^2 e^2\over 4\mn f_\pi^2}(\t_1+\t_3)
\Bigl[(d-1){\partial J_2[\omega,m^2]\over\partial \omega}
-(w^2+\half t){\partial J_0[\omega,m^2]\over\partial \omega}
\Bigl]+\hbox{crossed}\nonumber\\
T_b&=&-{g^2 e^2\over 2\mn f_\pi^2}\int_0^1\!dx\Bigl[
(\t_1+\t_3)(d+1){\partial J_2[x\omega,m^2]\over\partial x\omega}\nonumber\\&&
-\Bigl(x\omega^2(\t_2+\half \t_5)+(x^2w^2+\half x t)(\t_1+\t_3)\Bigr)
{\partial J_0[x\omega,m^2]\over\partial x\omega}\Bigr]+\hbox{crossed}\nonumber\\
T_c&=&-{g^2 e^2\over 2\mn f_\pi^2}\tau_3(\t_1+\t_3)
\Bigl(\omega J_0[\omega,m^2]+\Delta_\pi[m^2]\Bigr)+\hbox{crossed}\nonumber\\
T_d&=&{g^2 e^2\over 2\mn f_\pi^2}\tau_3(\t_1+\t_3) \int_0^1\!dx 
\Bigl(\omega J_0[x\omega,m^2]+\Delta_\pi[m^2]\Bigr)+\hbox{crossed}\nonumber\\
T_e&=&{g^2 e^2\over 2\mn f_\pi^2}(1-\tau_3)(\t_1+\t_3){1\over\omega}
\Big( J_2[\omega,m^2]-J_2[0,m^2]\Big)+\hbox{crossed}\nonumber\\
T_f&=&{g^2 e^2\over4 \mn f_\pi^2}\int_0^1\! dx\Bigl[
(\mu_v-\mu_s\tau_3)\Bigl( ((1-2x)\t_3+\t_1)\cos\theta- \t_2-(1-2x) \t_4\Bigr)
\omega J_0[x\omega,m^2]\nonumber\\&&
-(1-\tau_3)\Bigl(2(d+1)\t_1 \omega^{-1}J_2[x\omega,m^2]+
(2x(1-x)\t_1-\half(1-2x)\t_6)\omega J_0[x\omega,m^2]+\hbox{crossed}\nonumber\\
T_g&=&-{g^2 e^2\over 4\mn f_\pi^2}
\Bigl((\mu_v+\mu_s\tau_3)((\t_1+\t_3)\cos\theta-\t_2+\t_4-\t_5)+\half
(1+\tau_3)\t_6\Bigr)\,\omega
\int_0^1 \!dx J_0[x\omega,m^2]\nonumber\\&&\kern12cm+\hbox{crossed}\nonumber\\
T_h&=&-{g^2 e^2\over4\mn f_\pi^2}\int_0^1 \!dy \int_0^{1-y}\!dx\Bigl[
-(d+1)(d+3)\t_1{\partial J_2[\tilde\omega,m^2-xyt]\over\partial\tilde\omega}
\nonumber\\&&+\Big(\t_1(2(d+3)V(x,y)-t)+\omega^2\t_2((d+3)((d+5)xy-2x-2y)+2)+
\nonumber\\&&\qquad\qquad
((d+3)x-1)\omega^2(\t_6-\t_5)+(d+3)(1\!-\!x\!-\!y)\omega^2\t_4\Bigr)
{\partial J_0[\tilde\omega,m^2-xyt]\over\partial\tilde\omega}\nonumber\\
&&+\Bigl(-\t_1 V(x,y)^2-(x\t_6-x\t_5+(1\!-\!x\!-\!y)\t_4)V(x,y)\omega^2
\nonumber\\
&&\quad-((d+5)xy-x-y)(2V(x,y)\omega^2\t_2+(1\!-\!x\!-\!y)\omega^4\t_7)
+x y \,t \omega^2 \t_2\Big)
2{\partial J_0'[\tilde\omega,m^2-xyt]\over\partial\tilde\omega}
\nonumber\\
&&+\Bigl(\t_2 x y V(x,y)^2+\t_7xy(1\!-\!x\!-\!y)\omega^2V(x,y)\Bigr)
4\omega^2{\partial J_0''[\tilde\omega,m^2-xyt]\over\partial\tilde\omega}\Big]
+\hbox{crossed}\nonumber\\
T_i&=&{g^2 e^2\over4\mn f_\pi^2}\t_1\int_0^1\Bigl[(d^2-1)\Delta_\pi[m^2-x(1-x)t]
+2(1-2(d+1)x(1-x))t\Delta_\pi'[m^2-x(1-x)t]\nonumber\\&&
+4x^2(1-x)^2t^2\Delta_\pi''[m^2-x(1-x)t]\Big]\nonumber\\
T_j&=&{g^2 e^2\over\mn f_\pi^2}\int_0^1 \!dy \int_0^{1-y}\!dx\Bigl[
(d+1)\t_1\Delta_\pi[m^2-xyt]\nonumber\\&&
+\Bigl(-(2V(x,y)-t)\t_1+2(x+y-(d+3)xy)\omega^2\t_2\Bigr)
\Delta_\pi'[m^2-xyt]\nonumber\\&&+ (4V(x,y)-2t)xy\,\omega^2\t_2\,
\Delta_\pi''[m^2-xyt]\Bigr]\nonumber\\
T_k&=&-{g^2 e^2\over2\mn f_\pi^2}\t_1\int_0^1 \!dx\Bigl[
 (d-1)\Delta_\pi[m^2-x(1-x)t]
+(1-2x(1-x))t\, \Delta_\pi'[m^2-x(1-x)t]\Bigr]\nonumber\\
T_l&=&-{g^2 e^2\over\mn f_\pi^2}\t_1\Delta_\pi[m^2]\nonumber\\
T_m&=&-{g^2 e^2\over8\mn f_\pi^2}\t_1(3-\tau_3)(d-1)\Delta_\pi[m^2]\nonumber\\
T_n&=&-{e^2\over2f_\pi^2}\t_1\Bigl(8c_3-\smfrac 1 \mn (1+\tau_3)\Bigr)
\Delta_\pi[m^2]\nonumber\\
T_p&=&{e^2\over2f_\pi^2}\t_1\Bigl(16c_3-\smfrac 1 \mn (1+\tau_3)\Bigr)
\Delta_\pi[m^2]\nonumber\\
T_q&=&{8e^2\over f_\pi^2}\int_0^1 \!dy \int_0^{1-y}\!dx\Bigl[
-\half(\tilde c_2+(d+2)c_3)\t_1\Delta_\pi[m^2-xyt]\nonumber\\&&
+\Bigl((2 c_1\mpi^2-(\tilde c_2+c_3)\tilde\omega^2+V(x,y)c_3)\t_1+
\nonumber\\&&\qquad\qquad\qquad\qquad\qquad
(x y \tilde c_2+((d+4)xy-x-y) c_3)\omega^2\t_2\Bigr)\Delta_\pi'[m^2-xyt]\nonumber\\&&
+xy\omega^2\t_2\Bigl(-4c_1\mpi^2+2\tilde\omega^2(\tilde c_2+c_3)-2V(x,y)c_3\Bigr)
\Delta_\pi''[m^2-xyt]\Bigr]\nonumber\\
T_r&=&{2e^2\over f_\pi^2}\t_1\int_0^1\!dx\Bigl[(d c_3+\tilde{c}_2)
\Delta_\pi[m^2-x(1-x)t]\nonumber\\ &&
-(4c_1\mpi^2+2x(1-x)tc_3)\Delta_\pi'[m^2-x(1-x)t]\Bigr]\nonumber\\
T_s&=&{3g^2 e^2\over 8\mn f_\pi^2}(1+\tau_3)(d-1)\t_1\Delta_\pi[m^2]\nonumber\\
\end{eqnarray}
where the $c_i$ are low energy constants from ${\cal L}_{\N\pi}^{(2)}$, 
and $\tilde c_2=c_2-g^2/8\mn$.
The integrals $J_0[\omega,m^2]$, $J_2[\omega,m^2]$ and $\Delta_\pi[m^2]$ have
their usual meanings, prime denotes differentiation with respect to $m^2$,
$\tilde\omega=(1-x-y)\,\omega$ and 
\beq
V(x,y)=\tilde\omega^2+\half t(1-x-y+2xy).
\eeq
We have also introduced an extra tensor structure,
$\t_7= \sigbol\cdot(\qhat'\times \qhat)
\epsbol'\cdot\qhat\,\epsbol\cdot\qhat'.$  This is not independent;
$\t_7=\sin^2\theta \,\t_3+\cos\theta \,\t_5-\t_6$, but as it arises naturally 
in the calculations---and enters at too high an order in $\omega$ to affect
the polarisabilities---it is a useful notation.

The notation ``+ crossed" means that to every term, another is added  with
$\epsbol\leftrightarrow\epsbol'$, $\qvec\leftrightarrow-\qvec'$ and 
$\omega \leftrightarrow -\omega$.  Since the $\t_i$ are all either symmetric
or antisymmetric under 
this transformation, the net effect is to add a term with 
$\omega \leftrightarrow -\omega$ to the coefficients of $\t_1$ and $\t_2$,
and subtract such a term from the coefficients of $\t_3\ldots\t_7$.

The spin-dependent terms do not all look identical to the expressions given in 
ref.~\cite{kumar99}, but they are equivalent.


\begin{thebibliography}{99}

\bibitem{ber92} V. Bernard, N. Kaiser, J. Kambor and U.-G. Mei\ss ner, Nucl.\
Phys.\  {\bf B 388} 315 (1992).

\bibitem{pdg} D. E. Groom \etal,  European Physical Journal {\bf C15} 1 (2000). 

\bibitem{ber93} V. Bernard, N. Kaiser, A. Schmidt and  U.-G. Mei\ss ner,  
Phys.\ Lett.\ {\bf B 319} 269 (1993).

\bibitem{ber93a} V. Bernard, N. Kaiser, U.-G. Mei\ss ner and A. Schmidt, 
Z. Phys.\ {\bf A348} 317 (1993)

\bibitem{ber95} V. Bernard, N. Kaiser and U.-G. Mei\ss ner, 
Int.\ J. Mod.\ Phys.\ E {\bf 4} 193 (1995).

\bibitem{LGG} F. Low, Phys.\ Rev.\ {\bf 96} 1428 (1954); M. Gell-Mann and 
M. Goldberger, Phys.\ Rev.\ {\bf 96} 1433 (1954).
 
\bibitem{karl} I. Karliner, Phys.\ Rev.\ {\bf D7} 2717 (1973).

\bibitem{sand} A. M. Sandorfi, C. S. Whisnant and M. Khandaker, Phys.\ Rev.\
{\bf D50}  R6681 (1994).

\bibitem{drech98} D. Drechsel and G. Krein, Phys.\ Rev.\ {\bf D 58} 116009 
(1998).

\bibitem{tonn} J. Tonnison, A. M. Sandorfi, S. Hoblit and A. M. Nathan,
Phys.\ Rev.\ Lett.\ 80 4382 (1998).

\bibitem{wiss} F. Wissman, Talk at GDH-2000 symposium, Mainz, June 2000.

\bibitem{pedroni} P. Pedroni, Talk at GDH-2000 symposium, Mainz, June 2000.

\bibitem{mainz1} D. Drechsel, G. Krein and O. Hanstein, Phys.\ Lett.\ 
{\bf B 420} 248 (1998).

\bibitem{BGLMN} D. Babusci, G. Giordano, A. I. L'vov, G. Matone
and A. M. Nathan, Phys.\ Rev.\ {\bf C58} 1013 (1998).

\bibitem{mainz2} D. Drechsel,  M. Gorchtein, B. Pasquini and M. Vanderhaeghen,
Phys.\ Rev.\ {\bf C61} 015204 (2000).

\bibitem{ji00} X. Ji, C-W.\ Kao and J. Osborne, Phys.\ Rev.\ {\bf D 61} 
074003 (2000).

\bibitem{kumar99} K. B. V. Kumar, J. A. McGovern and M. C. Birse, 
{\tt hep-ph/9909442}.

\bibitem{gellas} G. C. Gellas, T. R. Hemmert and U.-G. Mei\ss ner, 
Phys.\ Rev.\ Lett.\ {\bf 85} 14 (2000).

\bibitem{kumar00} K. B. V. Kumar, J. A. McGovern and M. C. Birse, 
Phys.\ Lett.\ {\bf B 479} 167 (2000).

\bibitem{bab97} D. Babusci, G. Giordano and G. Mantone, Phys.\ Rev.\ {\bf C55}
R1645 (1997).

\bibitem{baranov}  P. S. Baranov \etal, Sov.\ J.\ Nucl.\ Phys.\ {\bf 21} 355 (1975).

\bibitem{baranov2} P. S. Baranov, A. L. L'vov, V. A. Petrun'kin  and L. N. Shtarkov,
Phys.\ Part.\ Nucl.\ {\bf 32} 376 (2001).

\bibitem{ziegler} A. Ziegler \etal, Phys.\ Lett.\ {\bf B 278} 34 (1992).

\bibitem{federspiel} F. J. Federspiel \etal,  Phys.\ Rev.\ Lett.\ {\bf 67} 1511 
(1991). 

\bibitem{hallin} E. L. Hallin \etal, Phys.\ Rev.\ {\bf C48} 1497 (1993).

\bibitem{macgibbon}B. E. MacGibbon \etal, Phys.\ Rev.\ {\bf C52} 2097 (1995).

\bibitem{deLeon} V. Olmos de Le\'on \etal, Eur. Phys. J. A {\bf 10} 207 (2001).

\bibitem{birse00} M. C. Birse, X. Ji, J. A. McGovern, {\tt nucl-th/0011054 }.

\bibitem{fettes00} N. Fettes and  U.-G. Mei\ss ner, Nucl.\ Phys.\ {\bf A 676} 
 311 (2000).
 
\bibitem{griess} H. W. Grie\ss hammer and G. Rupak, {\tt nucl-th/0012096}
\end{thebibliography}
\end{document}